# Statistical Reconstruction of arbitrary spin states of particles: root approach

Yu.I. Bogdanov

Institute of Physics and Technology, Russian Academy of Sciences,
Nakhimovskii pr. 34, 117218 Moscow, Russia[*]

**ABSTRACT**

A method of quantum tomography of arbitrary spin particle states is developed on the basis of the root approach. It is shown that the set of mutually complementary distributions of angular momentum projections can be naturally described by a set of basis functions based on the Kravchuk polynomials. The set of Kravchuk basis functions leads to a multi-parametric statistical distribution that generalizes the binomial distribution. In order to analyze a statistical inverse problem of quantum mechanics, we investigated the likelihood equation and the statistical properties of the obtained estimates. The conclusions of the analytical researches are approved by the results of numerical calculations.

## 1. INTRODUCTION

It is well known that statistical nature of quantum mechanics is its fundamental keystone. Any prediction of quantum theory is made by calculating the probability of some event. To compare experimental results with theoretical predictions one inevitably has to consider the problem of state vector reconstruction given experimental data. Naturally, the question is how should the state vector be reconstructed and what precision can be obtained. Until recently, this problem did not attract proper attention of scientists and only with the development of quantum informatics did it become relevant.

Traditionally, the methods of mathematical statistics and the mathematical apparatus of quantum mechanics have been considered separately to one another. Nonetheless, recent researches demonstrated that the two fields are closely interrelated. One of the goals of mathematical statistics was to develop a multiparametric statistical model that would allow stable reconstruction of unknown parameters by experimental data. In [1-5] it was shown that there is only one statistical model of such a kind, which is exactly the model that is considered in quantum mechanics. In view of these results, the quantum mechanical model has significant advantages compared to other statistical models. This special model is called rood approach. In terms of the root approach, the quantum mechanical psi function is a mathematical object of the analysis of data (both classical and quantum). The introduction of this function ensures an efficient solution of the inverse statistical problem.

An analysis of the multiparametric models shows that the conventional formalism of mathematical statistics, which does not use the notion of the psi function, is not so attractive as it would seem at first glance. Indeed, the classical objects of mathematical statistics are smooth parameterized families of densities determined up to one or two unknown parameters, which are the only quantities that should be estimated from the observational data, whereas the functional form of the density is assumed to be initially stringently defined (e.g., by Poissonian, Gaussian, etc., distributions). Attempts to estimate more complicated distributions, involving three, four, or more unknown parameters, run into severe calculational difficulties. In view of the ill-posedness of the corresponding inverse problems, the calculational difficulties, as well as those related to algorithm instabilities, rapidly become insurmountable with increasing dimension of a problem. The only model admitting stable solutions of multidimensional problems turns out to be the model based on the notion of the psi function

Methodologically, the root approach essentially differs also from other well-known methods for estimating quantum states. The latter techniques originate from extending the methods of classical tomography and classical statistics to quantum problems. Quantum analogues of the distribution density are the density matrix and the corresponding Wigner distribution function [6-8]. Therefore, by analogy with the methods of classical tomography, the methods developed so far have been aimed at reconstruction of the aforementioned objects (this explains the origin of the term "quantum tomography" [9]). Thus, in [10], a quantum tomography technique based on the Radon transformation of the Wigner function was proposed. The estimation of the quantum states by the least-squares method was considered in [11]. The strategy of a maximum likelihood was presented first in [12, 13]. A version of the maximum likelihood method in a form automatically ensuring the fulfillment of the basic conditions imposed on the density matrix (Hermitian character, nonnegative definiteness, and unitary trace) was presented in [14, 15]. Characteristic features of all these methods are rapidly increasing calculational complexity with increasing number of parameters to be estimated and, what is equally important, ill-posedness of the corresponding algorithms, which does not allow one to obtain correct stable solutions. The orientation toward the reconstruction of the density matrix overshadows the problem of estimation of a more fundamental object of the quantum theory, i.e., the state vector (the psi function). Formally, the states described by the psi function are particular cases of the states described by the density matrix. However, the fact that, in practice, the number of unknown parameters that should be estimated upon measurement of the density matrix is considerably greater than the corresponding number involved in the estimation of the psi function plays a key role in solution of statistical problems. Thus, in problems with the dimensionality $s$ the density matrix is determined by $s^2-1$ real-valued parameters, whereas, upon estimation of the psi function, one needs to determine only $2s-2$ coefficients, which, at large $s$, is a significantly smaller number. Equally important is the fact that the likelihood equation for the estimation of the psi function has a simple quasi-linear structure and can easily be solved numerically even when the number of parameters to be estimated amounts to hundreds or thousands. Note that a correct estimation of the density matrix by conventional methods runs into great difficulties even when the number of the parameters is of the order of ten. In the context of the root approach, the problem of estimation of the density matrix is reduced to the estimation of the states of pure components. The measurement results in combination with classical information about either the sources of particles or the environment allow one (at least, in principle) to reduce the study of the density matrix to the study of the mixture of components, each of which represents a pure state. It is necessary to keep in mind that the expansion of the mixture into pure components is ambiguous. However, the resultant density matrix is the same for any expansion within statistical fluctuations. If the classical information on the sources of particles is partially or totally unavailable, it is expedient to use a quasi-Bayesian algorithm of expanding the mixture into components, which was described in [3].

---

[*] E-mail: bogdanov@ftian.oivta.ru

Application of root approach to statistical analysis of experiments with biphoton field [16-20] proved the possibility to reconstruct the quantum states of 3- and 4-level optical systems with high precision that significantly exceeds the precision obtained in other works in the field.

Present work is based on applying root approach to the problem of spin state reconstruction. In our previous works [3,5] the simplest case of ½-spin particle state reconstruction was considered. In present work arbitrary spin particles are considered. It is shown that explicit analytical model of spin measures may be given and an effective numerical algorithm may be applied to its analysis. That is due to the close relation between quantum mechanical spin formalism and an orthogonal set of functions based on Kravchuk polynomials [21]. That important interrelation not only provides another example of applying functions of mathematical physics to quantum mechanics but is also important from statistical point of view. In particular, state of a $j$-spin particle may be set in correspondence to a set of discrete distributions that generalize binomial distribution with the number of experiments $N=2j$. By analogy to our previous works on quantum tomography [16-20], the theory of the spin quantum state estimate precision is developed. In particular, owing to the interconnection one may establish a correspondence between a $j$-spin particle and a set of discrete distributions that generalize binomial distribution with the number of experiments $N=2j$

A constructive criterion for estimating the precision of the correspondence between the theoretical state vector and its maximum likelihood estimate based on $\chi^2$-distribution is proposed. The validity of the criterion is tested by a numerical experiment.

## 2. LIKELIHOOD EQUATION FOR SPIN STATE RECONSTRUCTION

We will outline schematically the reconstruction of spin states of particles. Let $\psi_m^{(j)}$ be the amplitude of the probability to have the projection $m$ along the $z$ axis in the initial coordinate system (it is these quantities that should be estimated from the results of the measurements), $m=(j, j-1, \ldots, -j)$. Let $\widetilde{\psi}_m^{(j)}$ be the corresponding quantity in the rotated coordinate system. Upon measurement of the spin projection onto the $z'$ axis, the probability for this projection to be equal to $m$ is $\left|\widetilde{\psi}_m^{(j)}\right|^2$. The new and the old amplitudes are related to each other by the unitary transformation

$$\widetilde{\psi}_m^{(j)} = D_{mm'}^{*(j)} \psi_{m'}^{(j)} \qquad (1)$$

Here we assume summing on index $m'$

The matrix $D_{mm'}^{(j)}$ of finite rotations is a function of the Euler angles $D_{mm'}^{(j)}(\alpha, \beta, \gamma)$, where the angles $\alpha$ and $\beta$ coincide with the spherical coordinate angles of the $z'$ axis in the initial coordinate system $xyz$, with $\alpha = \varphi$ and $\beta = \theta$. The angle $\gamma$ corresponds to an additional rotation of the coordinate system about the $z'$ axis (upon measurement of the spin projection onto the $z'$ axis, this rotation is insignificant and, for definiteness, one can set $\gamma = 0$). The matrix of finite rotations is described in detail in [22]. Note that the transformation matrix in (1) corresponds to the inverse transformation with respect to the transformation considered in [22].

It is known that the unitary matrices of finite rotations $D_{mm'}^{(j)}(\alpha, \beta, \gamma)$ have the form

$$D_{mm'}^{(j)}(\alpha, \beta, \gamma) = \exp(im\gamma) d_{mm'}^j(\beta) \exp(im'\alpha) \qquad (2)$$

In [21] it was shown that the matrix elements $d_{mm'}^j(\beta)$ may be expressed by Kravchuk polynomials

$$d_{mm'}^j(\beta) = \frac{1}{d_n} \sqrt{\rho(x)} k_n^{(p)}(x, N) \qquad (3)$$

where $x = j - m'$, $n = j - m$, $N = 2j$, (4)

$$p = \sin^2(\beta/2), \quad \rho(x) = C_N^x p^x (1-p)^{N-x}, \qquad (5)$$

$$d_n = \sqrt{\frac{N!}{n!(N-n)!}(p(1-p))^n} \qquad (6)$$

The close interconnection between finite rotation matrices and an orthogonal set of Kravchuk functions allows one to consider spin as a natural generator of discrete binomial distributions. Let us derive a beam with the maximum spin projection $m = j$ on $z$-axis from the initial beam with Stein-Gerlach device. Then measuring the spin projection on $z'$-axis one gets binomial distribution $\rho(x) = C_N^x p^x (1-p)^{N-x}$, where $x$ is the number of "successes" in a series of $N$ independent experiments ($p$ – is the probability of success in a single experiment)

Equations (4)-(5) allow one to interconnect two different languages: spin-language and probability language. The probability of "success" $p = \sin^2(\beta/2)$ is defined by the angle $\beta$ between axes $z$ and $z'$, the number of trials $N$ is defined by the spin of the particles (from $N = 2j$ one may see that spin $j$ may be both integer and half-integer), the number of "successes" $x$ is defined by the difference between the maximum possible and the measured value of spin projection ($x = j - m'$).

From Eq. (3) one may see that the ordinary binomial distribution is a particular case that corresponds to the zero Kravchuk polynomial ($n = 0$). Thus, the ordinary binomial distribution is a generalized binomial distribution of zero order.

Similarly, in the aforementioned Stern-Gerlach scheme, by selecting other spin projections $m = j-1, j-2, ..., -j$ we generate distributions that may be called binomial distributions of order $n = 1, 2, .. N$. According to (3), these distributions correspond to proper orthogonal Kravchuk functions.

In general case, if the initial state is defined by a superposition of different spin projections, then one may consider the measurement of spin as a generator of some generalized binomial distribution. That distribution is a coherent superposition of distributions that correspond to different $n = 0, 1, .. N$.

According to the rules of quantum mechanics, any superposition may be obtained from the given initial state by some unitary transformation. In general case, however, that unitary transformation can not be expressed only in terms of rotations in space (one may need certain action on the spin from physical fields)

Let us consider amplitudes $\psi_m^{(j)}$ as a vector of state $c$ with the size $s = 2j+1$.

$$\begin{pmatrix} c_1 \\ c_2 \\ \vdots \\ c_s \end{pmatrix} = \begin{pmatrix} \psi_j^{(j)} \\ \psi_{j-1}^{(j)} \\ \vdots \\ \psi_{-j}^{(j)} \end{pmatrix}$$

Any direction of spin measurement is set in correspondence to a unique matrix $D_{mm'}^{(j)}(\alpha, \beta, \gamma)$ of size $s \times s$. A protocol that includes measuring the spin on $r$ different directions in space is described by some matrix $X$ of size $rs \times s$:

$$X = \begin{pmatrix} U_1 \\ U_2 \\ \vdots \\ U_r \end{pmatrix},$$

where each matrix $U_j$ ($j = 1, .., r$) corresponds to a certain matrix of finite rotations $D_{mm'}^{(j)}(\alpha, \beta, \gamma)$

Let us call matrix $X$ the instrumental matrix [16,17].

The complete set of measurements may be expressed as a set of $rs$ complex quantum amplitudes. In matrix representation one has

$$Xc = M, \qquad (7)$$

where the set of amplitudes is a column $M$ of size $rs$

$$M = \begin{pmatrix} \widetilde{c}_1 \\ \widetilde{c}_2 \\ \vdots \\ \widetilde{c}_r \end{pmatrix},$$

where $\widetilde{c}_j = U_j c$ ($j = 1, .., r$)

Each amplitude $M_\nu$ ($\nu = 1,..., rs$) defines some probability $p_\nu = |M_\nu|^2$ that corresponds to the proper spin projection on a given direction in space. Instead of the unknown amplitudes $M_\nu$ and probabilities $p_\nu = |M_\nu|^2$ we have a set of corresponding frequencies $k_\nu$ ($\nu = 1,..., rs$). The aim of quantum tomography is to estimate approximately the unknown state vector given a set of frequencies $k_\nu$. The considered problem may be solved with the maximum likelihood principle usage.

When analyzing the data of numerical experiments, we use the root estimator of quantum states [17]. This approach is designed specially for the analysis of mutually complementary measurements (in the sense of Bohr's complementarity principle). The advantage of this approach consists of the possibility of reconstructing states in a high-dimensional Hilbert space and reaching the accuracy of reconstruction of an unknown quantum state close to its fundamental limit.

Let us introduce the matrices with the elements defined by the following formulas:

$$I_{jl} = \sum_{i=1}^{rs} X_{ij}^* X_{il} \tag{8}$$

$$J_{jl} = \frac{r}{n_0} \sum_{i=1}^{rs} \frac{k_i}{p_i} X_{ij}^* X_{il} \quad j,l = 1,...,s \tag{9}$$

where $n_0 = \sum_{i=1}^{rs} k_i$ - is the total number of particles that were conducted to spin measurement

The matrix $I$ is determined from the experimental protocol and, thus, is known *a priori* (before the experiment). This is the Hermitian matrix of Fisher's information. The matrix $J$ is determined by the experimental values of $k_i$ and by the unknown probabilities $p_i$. This is the empirical matrix of Fisher's information.

In terms of these matrices, in complete analogy to [16,17], in our case, the likelihood equation has the form

$$I^{-1}Jc = c \tag{10}$$

This is a nonlinear equation, because p$i$ depends on the unknown state vector $c$. Because of the simple quasi-linear structure, this equation can easily be solved by the iteration method [1,2].

### 3. STATISTICAL FLUCTUATIONS OF THE STATE VECTOR

The fluctuations of the quantum state in a normally functioning quantum information system should be within a certain range defined by the statistical theory. The present section is devoted to this problem. The expressed method is similar to the method expressed in our work [17]

The practical significance of accounting for statistical fluctuations in a quantum system relates to developing methods of estimation and control of the precision and stability of a quantum information system evolution, as well as methods of detecting external interception (Eve's attack on the quantum channel between Alice and Bob).

The estimate of the state vector $c$, obtained by the maximum-likelihood principle, differs from the exact state vector $c^{(0)}$ by the small random values $\delta c = c - c^{(0)}$. Let us consider the statistical properties of the fluctuation vector $\delta c$ by expansion of the log-likelihood function near a stationary point:

$$\delta \ln L = -\left[\frac{1}{2}\left(K_{sj}\delta c_s \delta c_j + K_{sj}^*\delta c_s^* \delta c_j^*\right) + I_{sj}\delta c_s^* \delta c_j\right] \quad (11)$$

Together with the Hermitian matrix of the Fisher information $I$, Eq. (8), we define the symmetric Fisher information matrix $K$, whose elements are defined by the following equation:

$$K_{sj} = \sum_v \frac{p_v}{M_v^2} X_{vs} X_{vj} \quad (12)$$

In the general case, $K$ is a complex symmetric non-Hermitian matrix.

The complex fluctuation vector $\delta c$ is conveniently represented by a real vector of double length. After extracting the real and imaginary parts of the fluctuation vector $\delta c_j = \delta c_j^{(1)} + i\delta c_j^{(2)}$ we transfer from the complex vector $\delta c$ to the real one $\delta \xi$:

$$\delta c = \begin{pmatrix} \delta c_1 \\ \delta c_2 \\ \vdots \\ \delta c_s \end{pmatrix} \rightarrow \delta \xi = \begin{pmatrix} \delta c_1^{(1)} \\ \vdots \\ \delta c_s^{(1)} \\ \delta c_1^{(2)} \\ \vdots \\ \delta c_s^{(2)} \end{pmatrix} \quad (13)$$

This transition provides us with a $2s$-component real vector instead of a $s$-component complex vector.

In the new representation, Eq. (11), becomes

$$\delta \ln L = -H_{sj}\delta \xi_s \delta \xi_j = -\langle \delta \xi | H | \delta \xi \rangle \quad (14)$$

where matrix $H$ is the "complete information matrix" possessing the following block form:

$$H = \begin{pmatrix} \text{Re}(I+K) & -\text{Im}(I+K) \\ \text{Im}(I-K) & \text{Re}(I-K) \end{pmatrix} \quad (15)$$

The matrix $H$ is real and symmetric. It is of double dimension, respectively, to the matrices $I$ and $K$ ($I$ and $K$ are $s \times s$ matrices, while $H$ is $2s \times 2s$).

Knowledge of the numerical dependence of statistical fluctuations allows one to estimate distributions of various statistical characteristics. The important information criterion that specifies the general possible level of statistical fluctuations in a quantum information system is the $\chi^2$ - criterion.

Similarly to Eq. (13), let us introduce the transformation of a complex state vector to a real vector of double length:

$$c = \begin{pmatrix} c_1 \\ c_2 \\ \vdots \\ c_s \end{pmatrix} \rightarrow \xi = \begin{pmatrix} c_1^{(1)} \\ \vdots \\ c_s^{(1)} \\ c_1^{(2)} \\ \vdots \\ c_s^{(2)} \end{pmatrix} \quad (16)$$

It can be shown that the information carried by a state vector is equal to the doubled total number $r$ of mutually-complementary directions that were introduced for spin measurement

$$\langle \xi | H | \xi \rangle = 2r \tag{17}$$

Then, the $\chi^2$ - criterion can be expressed in a form:

$$\frac{\langle \delta\xi | H | \delta\xi \rangle}{\langle \xi | H | \xi \rangle} \sim \frac{\chi^2(2s-2)}{4n_0}, \tag{18}$$

where $s$ is the Hilbert space dimension, $n_0$ is the total number of observations.

Relation (18) describes the distribution of relative information fluctuations. It shows that the relative information uncertainty of a quantum state decreases with the number of observations as $1/n_0$.

The left-hand side of Eq. (18), which describes the level of state vector information fluctuations, is a random value, based on the $\chi^2$ - distribution with $2s-2$ degrees of freedom.

Let us note that a vector that consists of $s$ complex numbers is defined by $2s$ real numbers. The reduce of the number of degrees of freedom by two (from $2s$ to $2s-2$) is done by normalizing state vector length for unit and the arbitrary phase of state vector (global gauge invariance). In some cases it is convenient not to define the normalization condition for state vector a priori (than the normalization condition is defined by the total intensity of all the considered quantum processes and has to be estimated itself). If the state vector is not assumed to be normalized a priori, then for $\chi^2$ distribution we will have $2s-1$ degrees of freedom rather than $2s-2$ [17].

The mean value of relative information fluctuations is

$$\frac{\overline{\langle \delta\xi | H | \delta\xi \rangle}}{\langle \xi | H | \xi \rangle} = \frac{2s-2}{4n_0} \tag{19}$$

The information fidelity may be introduced as a measure of correspondence between the theoretical state vector and its estimate:

$$F_H = 1 - \frac{\langle \delta\xi | H | \delta\xi \rangle}{\langle \xi | H | \xi \rangle} \tag{20}$$

Correspondingly, the value $1-F_H$ is the information loss.

The convenience of $F_H$ relies on its simpler statistical properties compared to the conventional fidelity $F = |\langle c_{theor} | c_{exp} \rangle|^2$. For a system where statistical fluctuations dominate, fidelity is a random value, based on the $\chi^2$ - distribution:

$$F_H = 1 - \frac{\chi^2(2s-2)}{4n_0} \tag{21}$$

where $\chi^2(2s-2)$ is a random value of $\chi^2$ - type with $(2s-2)$ degrees of freedom.

Information fidelity asymptotically tends to unity when the sample size is growing up. Complementary to statistical fluctuations noise (that is connected with instrumental errors) leads to a decrease in the informational fidelity level compared to the theoretical level (21).

The validity of the analytical expression (18) is justified by the results of numerical modeling and observed data. As an example on fig1 the results of numerical experiment are given that prove the $\chi^2$ criterion in form (18). The demonstrated data corresponds to 300 numerical experiments, where for each experiment the state of particles with spin $j = 10$ ($s = 2j + 1 = 21$) was estimated by measuring $n_0 = 30000$ representatives of quantum statistical ensemble (10000 measurements in each of the three mutually-orthogonal directions in space).

Fig. 1 Spin States Reconstruction: the results of numerical modeling

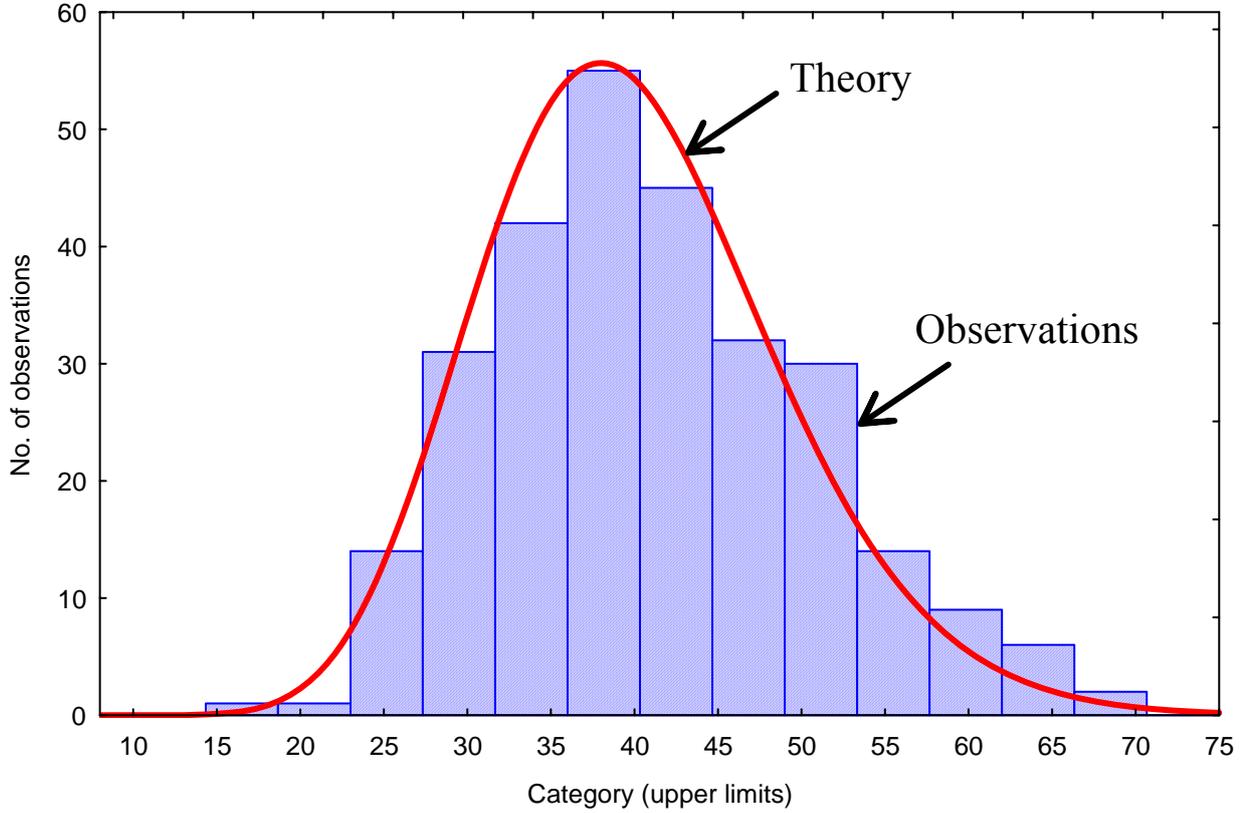

As it is evident from the numerical calculations similar results may be obtained for other spin particles ($j = 1/2, 1, 3/2...$ etc.).

## 4. CONCLUSIONS

Let us formulate main results of the work:
1. A method of quantum tomography of arbitrary spin particle states is developed on the basis of the root approach. It is shown that one may develop an explicit analytical model of spin measurements and a convenient and efficient algorithm of its numerical analysis.
2. It is shown that the set of mutually complementary distributions of angular momentum projections can be naturally described by a set of basis functions based on the Kravchuk polynomials. The set of Kravchuk basis functions leads to a multi-parametric statistical distribution that generalizes the binomial distribution. It is shown that one may establish a correspondence between a set of discrete distributions that generalize binomial distribution with the number of trials $N = 2j$ and a $\dot{j}$-spin particle
3. In order to analyze a statistical inverse problem of quantum mechanics, we investigated the likelihood equation and the statistical properties of the obtained estimates. A constructive criterion based on $\chi^2$-distribution for estimating precision of correspondence between theoretical state vector and its maximum likelihood estimate is proposed.
4. The conclusions of the analytical researches are approved by the results of numerical calculations.

Present work was supported by a grant of the Federal agency of science and innovations on subject "Quantum information methods" conducted by a scientific group with the head academician K.A.Valiev.